\documentclass{llncs}

\usepackage{verbatim}
\usepackage{url}
\usepackage{epsfig}

\begin{document}

\title{Say Hi to Eliza \subtitle{An Embodied Conversational Agent on the Web}}

\author{Gerard Llorach\inst{1,}\inst{2,}\inst{3} \and Josep Blat\inst{1}}
\authorrunning{Gerard Llorach et al.}

\institute{
Interactive Technologies Group, Universitat Pompeu Fabra, Barcelona, Spain\\
\email{\{gerard.llorach, josep.blat\}@upf.edu\\}
\and Medizinische Physik and Cluster of Excellence ‘Hearing4all', Universit\"at Oldenburg\\
\and H\"orzentrum Oldenburg GmbH, Oldenburg, Germany
}

\maketitle

\begin{abstract}
The creation and support of Embodied Conversational Agents (ECAs) has been quite challenging, as features required might not be straight-forward to implement and to integrate in a single application. Furthermore, ECAs as desktop applications present drawbacks for both developers and users; the former have to develop for each device and operating system and the latter must install additional software, limiting their widespread use. In this paper we demonstrate how recent advances in web technologies show promising steps towards capable web-based ECAs, through some off-the-shelf technologies, in particular, the Web Speech API, Web Audio API, WebGL and Web Workers. We describe their integration into a simple fully functional web-based 3D ECA accessible from any modern device, with special attention to our novel work in the creation and support of the embodiment aspects.

\end{abstract}

\begin{figure}
\epsfysize=5cm 
\hspace{0.8cm}
\epsfbox{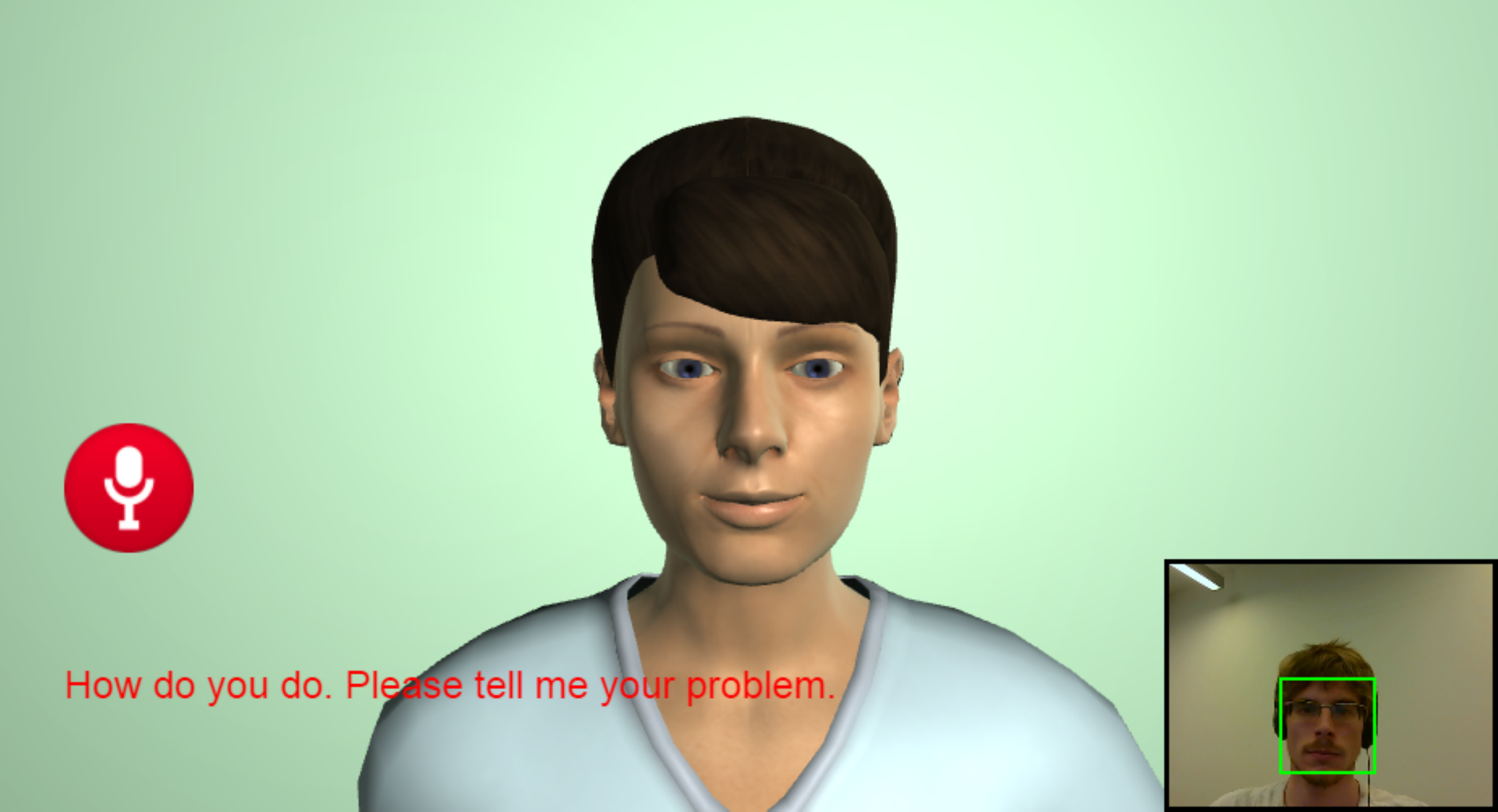}

\caption[]{Interface of the web application}
\end{figure}

\begin{tiny}
The final publication is available at Springer via http://dx.doi.org/10.1007/978-3-319-67401-8\_34
\end{tiny}

\section{Introduction}

We present an implementation where our main contribution is the support of the 3D embodiment and the integration of web technologies. We demonstrate that a 3D ECA in the web browser is feasible using the right tools and libraries, with the work of one or two experts during two weeks. In addition, our system is open and standars based, so that better/alternative artificial intelligence or other components, can be easily connected. Indeed, our web-based ECA uses the following web/open source components:

\begin{enumerate}
\item \textbf{Listening and Speaking}: speech recognition and synthesis with the Web Speech API.
\item \textbf{Understanding, thinking and replying}: ELIZA \cite{eliza} as artificial conversational entity; rule-based nonverbal behaviors described in \cite{lookmeintheeyes} for gaze and head motions.
\item \textbf{Embodiment}: creation of virtual characters with Makehuman and Blender; support, real-time rendering and animations with WebGLStudio \cite{webglstudio}; facial tracking with javascript libraries (jsfeat) \cite{jsfeat} within a Web Worker allowing the agent to follow the user with the gaze.
\end{enumerate}

Currently there are no WebGL-based ECAs with 3D virtual characters with the advanced features we present to the best of our knowledge.

\section{Related Work}

RAG LiteBody \cite{dtask} was one of the first web implementations but the ECAs only allowed the user to choose from a set of sentences as input and the embodiment was 2D and based on Adobe Flash. A first WebGL implementation of a talking head can be seen in \cite{lucia}, but the facial features need to be processed by the server and it is not interactive. The company Existor \cite{existor} with products such as Cleverbot, Cleverscript and Evie, is one of the few that supports web-based ECAs, although they are more specialized on the creation of the chatbot system and use off-the-shelf web components for speech processing. They embody the agent through 2D video-realistic facial expressions synthesis by means of morphing with Adobe Flash plugins. Our approach does not need any plugins, it is based on open source components and allows the users and researchers to modify any components and to create their own virtual characters.

\section{Components}

\subsubsection{Speech Recognition, Speech Synthesis, Dialogue System}
The Web Speech API permits to use local OS services and external services by URI. In the Chrome browser, Google services are the default configuration, and provide real-time, incremental speech recognition in several languages and dialects with high accuracy as well as speech synthesis through a few lines of code.

For demonstration purposes we used one of the first chatbots: ELIZA from the mid 60s. We integrated the chatbot as a script on the web client \cite{elizajs}, adding some nonverbal behavior such as shaking the head when negation words (no, not, n't) are spoken by the agent and head nods while listening and speaking \cite{lookmeintheeyes}. It is important to note that more sophisticated dialogue systems can be as easily connected, and provide a more conversationally capable ECAs.

\subsubsection{Embodiment}

Among the tools to create 3D humanoid characters, Makehuman, Autodesk Character Generator, Poser, Mixamo (Fuse) and Daz Studio, we used Makehuman to create virtual characters and Blender to add the blend shapes and optimize the models.

Applications such as Unity3D and Unreal Engine permit to rapidly create, visualize and export 3D scenes and games with little coding and compatibility problems for different OSs. However, web plugins are becoming unsupported by some browsers, in particular the Unity Web Player by Chrome. 
 We chose WebGLStudio, a 3D scene editor and game engine, as it has better tools and components to easily to integrate virtual characters than other web engines such as PlayCanvas \cite{playcanvas} and Clara.io \cite{claraio} and it is supported in Chrome.

Our implementation supports some basic BML commands such as gaze and head nods in WebGLStudio; uses a web-based system which automatically generates facial expressions based on the two values of valence and arousal as proposed in [12] and a web-based lip-syncing implementation proposed in [13]. Only eight blend shapes are needed for facial expressions, where three of them are used for the lip-sync, which is quite cost-effective when creating new characters.

We used a facial tracking library \cite{jsfeat} and implemented it in Web Workers to extract the position of the user's face relative to the camera so that the ECA directs its gaze at the user. Facial expression analysis libraries and algorithms could also be implemented using such Web Workers, so that the nonverbal behavior of the user could be extracted and used in the dialogue.

\section{Results, Discussion and Future Work}
A novel web-based ECA was implemented successfully, fulfilling all basic requirements. The timing performance of each component  was tested over 100 interactions on a PC (Windows 8 x64 2.50GHz, NVIDIA GeForce GT 750M) with Google Chrome v56.0.2924.87 and a internet connection of 45Mbps with the following results: SST $mean= 166.67ms$ ($sd=86.47$); TTS $mean=233.01ms$ ($sd=251.39$); ElizaAI $mean=3.72ms$ ($sd=1.88$); Total processing time $mean=392.66ms$ ($sd=254.09$). Thus, the processing time of an interaction with the system (STT, TTS and AI) is less than a second, an acceptable pause in natural human conversations \cite{sacks}. The system was developed in two weeks of work, using WebGL, BML and facial tracking libraries.

Open web-based ECAs such as ours will allow researchers to carry out large user studies more easily and  possibly in a more standardized way, a contribution to advance in the ECAs research field.

\section{Acknowledgments}
This research has been partially funded by the Spanish Ministry of Economy and Competitiveness (RESET TIN2014-53199-C3-3-R), by the DFG research grant FOR1732 and by the European Commission under the contract number H2020-645012-RIA (KRISTINA) and under the  the Marie Sklodowska-Curie grant agreement No 675324 (ENRICH). Special thanks to Volker Hohmann and Sergio Sagayo for revisions and counseling and to Javi Agenjo for developing WebGLStudio and helping out with all the technical challenges.

\end{document}